%% file: main.tex
\documentclass{article}

\usepackage{spconf,amsmath,graphicx,amssymb}

\usepackage{wrapfig}
\usepackage{graphicx}
\usepackage{subfigure}
\usepackage{float}
\usepackage{adjustbox}
\usepackage[justification=centering]{caption}
\usepackage{tabularx}
\usepackage{multirow}
\usepackage[belowskip=-15pt,aboveskip=0pt]{caption}
\usepackage{array}
\newcolumntype{P}[1]{>{\centering\arraybackslash}p{#1}}
\newcolumntype{M}[1]{>{\centering\arraybackslash}m{#1}}
\newcommand{\PreserveBackslash}[1]{\let\temp=\\#1\let\\=\temp}
\newcolumntype{R}[1]{>{\PreserveBackslash\raggedleft}p{#1}}
\usepackage{tabulary,booktabs}
\usepackage{enumitem}
\usepackage{lipsum}
\usepackage{blindtext}
\usepackage[linesnumbered, algo2e, ruled,norelsize]{algorithm2e}
\usepackage{xcolor}
\usepackage{cite}

\usepackage{hyperref}

\SetCommentSty{mycommfont}
\title{Learning-based lossless compression of 3D point cloud geometry}
\name{Dat Thanh Nguyen, Maurice Quach, Giuseppe Valenzise, Pierre Duhamel}
\address{Universit\'e Paris-Saclay, CNRS, CentraleSup\'elec, Laboratoire des Signaux et Syst\`emes\\
91190 Gif-sur-Yvette, France}
\begin{document}
%\ninept
%
\maketitle
\begin{abstract}
\input{abstract}
\end{abstract}
\begin{keywords}
Point Cloud Compression, Deep Learning, G-PCC, context model.
\end{keywords}
\section{Introduction}
\label{sec:intro}
\input{introduction}
\section{Related work}
\label{sec:stateoftheart}
\input{stateOfTheArt}

\section{Proposed method}
\label{proposedmethod}
\input{proposed}

\section{Experimental Results}
\label{performanceeval}
\input{perform}

\section{Conclusions}
\label{conclusion}

\input{conclusion}

% Below is an example of how to insert images. Delete the ``\vspace'' line,
% uncomment the preceding line ``\centerline...'' and replace ``imageX.ps''
% with a suitable PostScript file name.
% -------------------------------------------------------------------------

% \vfill\pagebreak

%\section{REFERENCES}
%\label{sec:refs}

\bibliographystyle{IEEEbib}
\bibliography{refs.bib}

% \newpage
% \section{Ablation Study}
%\input{AblationStudy}
\end{document}

%% file: abstract.tex
This paper presents a learning-based, lossless compression method for static point cloud geometry, based on context-adaptive arithmetic coding.  Unlike most existing methods working in the octree domain, our encoder operates in a \textit{hybrid} mode, mixing octree and voxel-based coding. We adaptively partition the point cloud into multi-resolution voxel blocks according to the point cloud structure, and use octree to signal the partitioning. On the one hand, octree representation can eliminate the sparsity in the point cloud. On the other hand, in the voxel domain, convolutions can be naturally expressed, and geometric information (i.e., planes, surfaces, etc.) is explicitly processed by a neural network.  Our context model benefits from these properties and learns a probability distribution of the voxels using a deep convolutional neural network with masked filters, called VoxelDNN. Experiments show that our method outperforms the state-of-the-art MPEG G-PCC standard with average rate savings of 28\% on a diverse set of point clouds from the Microsoft Voxelized Upper Bodies (MVUB) and MPEG. The implementation is available at \url{https://github.com/Weafre/VoxelDNN}.

%where the geometry information (i.e. curve, plane...) is completely preserved and convolution can be naturally expressed.
%Most existing lossless point cloud geometry compression solutions are implemented in the octree domain which efficiently deals with the sparsity of point clouds.
%Point Cloud Compression (PCC) is attracting more and more attention from the research community because of its significant impact on immersive applications (e.g. augmented reality and virtual reality) as well as autonomous vehicles.
%In addition, we proposed a second solution consisting of a scalable point cloud coding based on an existing lossy coder using auto-encoder architecture.

%The proposed solution partitions the point clouds into non-empty fixed-size blocks and selects the best coding approach between encoding the whole block as one and further partitioning before actually encoding each voxel to deal with the high sparsity blocks. 

%We trained a model with conditional probabilities generated by a deep neural network. 
% we believe previously encoded voxel will give a valuable information of the current voxel

%% file: introduction.tex
\par Recent visual capturing technology has enabled 3D scenes to be captured and stored in the form of Point Clouds (PCs). PCs are now becoming the preferred data structure in many applications (e.g., VR/AR, autonomous vehicle, cultural heritage), resulting in a massive demand for efficient Point Cloud Compression (PCC) methods. 
%Typical PCs contain millions of points and each point is represented by spatial coordinates and attributes (e.g. color, reflectance, etc.) which require a huge amount of storage.
\par The Moving Picture Expert Group has been developing two PCC standards \cite{8571288,jang2019video,graziosi2020overview}: Video-based PCC (V-PCC) and Geometry-based PCC (G-PCC). V-PCC focuses on dynamic point clouds and is based on 3D-to-2D projections. On the other hand, G-PCC targets static content and encodes the point clouds directly in 3D space. In G-PCC, the geometry and attribute information are independently encoded. However, the geometry must be available before filling point clouds with attributes. Therefore, having an efficient lossless geometry coding is fundamental for efficient PCC. A typical point cloud compression scheme consists in pre-quantizing the geometric coordinates using voxelization. In this paper, we also adopt this approach, which is particularly suited for dense point clouds. After voxelization, the point cloud geometry can be represented either directly in the voxel domain, or using an octree spatial decomposition.

\par In this work, we propose a deep-learning-based method for lossless compression of voxelized point cloud geometry. Using a masked 3D convolutional network, our approach (named VoxelDNN) first learns  the distribution of a voxel given all previously decoded ones. This conditional distribution is then used to model the context of a context-based arithmetic coder. In addition, we reduce point cloud sparsity by adaptively partitioning the PC.  We demonstrate experimentally that the proposed solution outperforms the MPEG G-PCC solution in terms of bits per occupied voxel with average rate savings of $28\%$ on all test datasets. 

The rest of the paper is structured as follows: Section \ref{sec:stateoftheart} reviews the related work; the proposed method is described in Section \ref{proposedmethod}; Section \ref{performanceeval} presents the experimental results; and finally Section \ref{conclusion} concludes the paper.
%The volume is divided vertically and horizontally into pre-defined size cubes and each cube is called a voxel. If a voxel contains at least 1 point, it is called an occupied voxel. Usually, less than 1\% of voxels are occupied.  Octree representation \cite{meagher1982geometric} is obtained by analyzing the geometry of the voxelized point cloud. Assuming that the point cloud is mapped into a volume of $D \times D \times D$ voxels, the volume is recursively split into eight sub-cubes until the desired precision is achieved. Then, occupied blocks are marked by bit 1 and empty blocks are marked by bit 0. Consequently, at each level, the generated 8 bits represents the occupancy state of an octree node (octant).
%There are many existing lossless methods to encode point cloud geometry including the geometry coder of G-PCC.

%\begin{figure}[htb]

%\begin{minipage}[b]{1.0\linewidth}
%  \centering
%  \centerline{\includegraphics[width=\linewidth]{figures_pcc/Screenshot 2020-09-12 at 15.09.10.png}}
%  \vspace{2.0cm}
%\caption{Octree partitioning \cite{graziosi2020overview}}
%\label{fig:octreepartitioning}
%\end{minipage}
%\end{figure}

%% file: stateoftheart.tex
% \par Our work draws ideas from recently proposed deep generative models and learning-based point cloud compression methods. 
\par Most existing point cloud geometry compression methods, including MPEG G-PCC test models,  represent and encode occupied voxels using octrees \cite{garcia2017context, garcia2018intra, garcia2019geometry} or local approximations called ``triangle soups''~\cite{dricot2019adaptive}. Recently, the authors of \cite{garcia2019geometry} proposed an intra-frame method called P(PNI), which builds a reference octree by propagating the parent octet to all children nodes, thus providing 255 contexts to encode the current octant. A frequency table of size $255 \times 255$ is built to encode the octree and needs to be transmitted to decoder. A drawback of this octree representation is that, at the first levels of the tree, it produces ``blocky'' scenes, and geometry information of point clouds (i.e., curve, plane) is lost. Instead, in this paper we work in the \textit{hybrid domain} to exploit the geometry information. In addition, our method predicts voxel distributions in a sequential manner at the decoder side, thus avoiding the extra cost of transmitting large frequency tables.

\par Our work draws inspiration from the recent advances in deep generative models for 2D images. The goal of generative models is to learn the data distribution, which can be used for a variety of tasks,  with image generation being probably the most popular~\cite{goodfellow2016deep}. Among the various classes of generative models, we consider  methods able to explicitly estimate data likelihood, such as the PixelCNN model \cite{oord2016pixel,salimans2017pixelcnn++}.  
% and learning-based image compression \cite{balle2016end,mentzer2018conditional}.  
Specifically, these approaches factorize the likelihood of a picture by modeling the conditional distribution of a given pixel's color given all previously generated pixels. 
% These conditional distributions only depend on the possible pixel values with respect to the scanned context, which imposes some \textit{causality} constraint. 
PixelCNN models the distribution using a neural network. The causality constraint is enforced using masked filters in each convolutional layer. 
% The proposed lossy image compression in \cite{mentzer2018conditional} also uses masking filters to estimate the entropy $H$ of a latent representation as the measure of rate $R$ . 
Recently, this approach has been employed in image compression to yield accurate and learnable entropy models~\cite{mentzer2018conditional}.
This paper extends the generative model and masking filters to 3D point cloud geometry compression.

\par Inspired by the success in learning-based image compression, deep learning has been recently adopted in point cloud coding methods  \cite{huang2020octsqueeze,8954537,9191021,quach2019learning,quach2020improved}. The proposed methods in \cite{quach2019learning,quach2020improved}  encode each $64 \times 64 \times 64$ sub-block of PC using a 3D convolutional auto-encoder. In contrast, in this paper we losslessly encode the voxels by directly learning the distribution of each voxel from its 3D context. We also offer block-based coding, which was  successful in traditional image and video coding.

%\par Our compression method in this report learns the distribution of a voxel from the context in 3D space which is all the previously scanned voxels using masked 3D Voxel CNN model. 
% Slicing 3D Boolean array G(x, y, z) of size N x N x N along with one of each axis and encoding each of them using JBIG
% Encoder encodes input from high-dimensional space to low-dimensional space while the decoder trying to reconstruct input.
%Octree-based PCC methods efficiently eliminate the sparsity of the point cloud and have scalability.

%% file: proposed.tex
\subsection{Definitions}
\label{definition}
 A point cloud voxelized over a $2^n \times 2^n \times 2^n$ grid is known as an $n$-bit depth PC, which can be represented by an $n$ level octree. In this work, we represent point cloud geometry in a hybrid manner: both in octree and voxel domain. We coarsely partition an $n$-depth point cloud up to  level $n-6$. As a result, we obtain a $n-6$ level octree and a number of non-empty binary blocks $v$ of size $2^6 \times 2^6 \times 2^6$ voxels, which we refer to as resolution $d=64$ or simply block 64 in the following. Blocks 64 can be further partitioned at resolution $d=\{32, 16, 8, 4\}$ as detailed in Section~\ref{ssec:multires}. At this point, we encode the blocks using our proposed encoder in the voxel domain (Section~\ref{ssec:voxelDNN}). The high-level octree, as well as the depth of each block, are converted to bytes and signaled to the decoder as side information. We index all voxels in block $v$ at resolution $d$ from $1$ to ${d^3}$ in raster scan order with:
 \begin{equation}
    v_i= 
    \begin{cases} 
    1, \quad \text{if $i^{th}$ voxel is occupied}\\
    0, \quad \text{otherwise}.
    \end{cases}
\label{focalloss}
\end{equation}
\par 
 %A point cloud is a set of points in a 3D space. Each point is associated with coordinate information together with attributes (e.g.color, normals, etc.). 
%\input{scalable}
 %The location of points in 3D space can be quantized at a given precision. Usually, 3D space is divided into a grid of size $D \times D \times D$ where D is a power of 2 and each sub-grid can contains multiple points which are mapped to the sub-grid center coordinates. The process of creating the grid-based coordinates is known as voxelization when each sub-grid is called as voxel.
 %Figure \ref{fig:partition} illustrates the process creating octree. 
%The geometry information of voxelized point cloud could be represented in another structure called octree. As mentioned before, a voxelized point cloud is recursively divided into 8 sub-cubes and we use 8 bits to indicate whether a cube is occupied or not.
%\begin{figure}[hbt]
%\centering
%\captionsetup{justification=centering}
%\includegraphics[width=\linewidth]{figures_pcc/partitioning.png}
%\caption{Voxel parititioning of a depth 8 point cloud into 2 level octree and non-empty block of size 64x64x64 indicated by green color.}
%\label{fig:partition}
%\end{figure}
\input{voxelDNN}

%% file: voxelDNN.tex
%\par Point cloud data structure, voxel domain, octree domain
\subsection{VoxelDNN}\label{ssec:voxelDNN}

\begin{figure}
\captionsetup{singlelinecheck = false, format= hang, justification=raggedright, font=small, labelsep=space}
\begin{minipage}[b]{.45\linewidth}
  \centering
  \centerline{\includegraphics[width=0.65\linewidth]{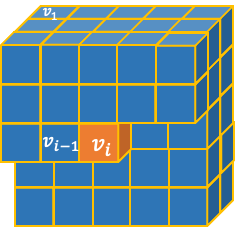}}
%  \vspace{1.5cm}
  \centerline{(a) 3D voxel context}\medskip
\end{minipage}
\hfill
\begin{minipage}[b]{0.45\linewidth}
\label{sfig:typeA}
  \centering
  \centerline{\includegraphics[width=0.85\linewidth]{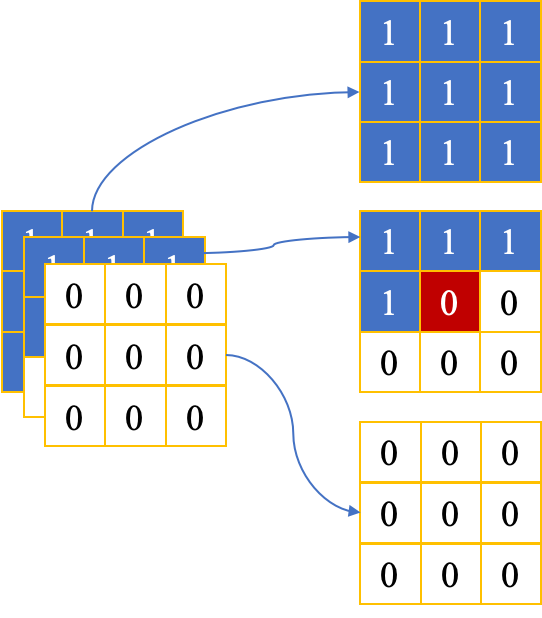}}
%  \vspace{1.5cm}
  \centerline{(b) 3D type A mask }\medskip
\end{minipage}
\caption{(a): Example 3D context in a $5 \times 5 \times 5$ block. Previously scanned elements are in blue. (b): $3 \times 3 \times 3$ 3D type A mask. Type B mask is obtained by changing center position (marked red) to 1. }
\label{fig:context}
\end{figure}
\begin{figure}[tb]
\captionsetup{singlelinecheck = false, format= hang, justification=raggedright, font=small, labelsep=space}
\centering
\includegraphics[width=0.9\linewidth]{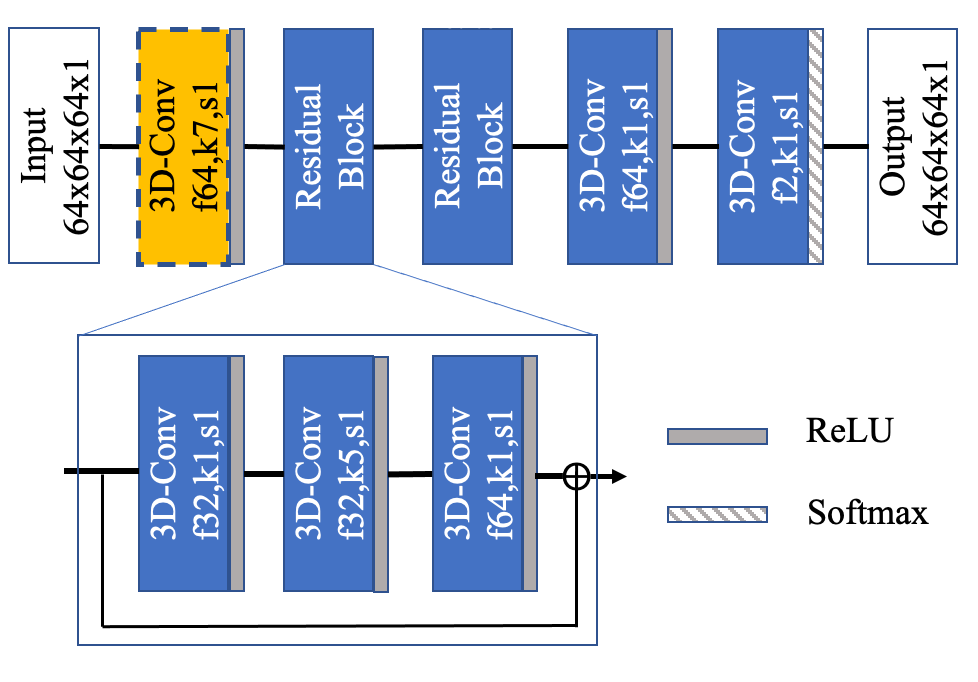}
\caption{VoxelDNN architecture. A type A mask is applied in the first layer (dashed borders) and type B masks afterwards. `f64,k7,s1' stands for 64 filters, kernel size 7 and stride 1. }
\label{fig:Networkarchitecture}
\end{figure}
\setlength{\textfloatsep}{20pt}% Remove \textfloatsep

\input{encoder_algrth}

% \subsubsection{Voxel context model}
\par Our method encodes the voxelized point cloud losslessly using context-adaptive binary arithmetic coding. Specifically, we focus on estimating accurately a probability model $p(v)$  for the occupancy of a block $v$ composed by $d \times d \times d$ voxels. We factorize the joint distribution $p(v)$ as a product of conditional distributions $p(v_i|v_{i-1}, \ldots, v_1)$ over the voxel volume: 
\begin{equation}
    p(v)= \underset{i=1 }{\overset{d^3}{\Pi}}p(v_i|v_{i-1},v_{i-2},\ldots,v_{1}).
    \label{eq:p(v)}
\end{equation}
Each term $p(v_i|v_{i-1}, \ldots, v_1)$ above is the probability of the voxel $v_{i}$ being occupied given all previous voxels,  referred to as a context. Figure \ref{fig:context}(a) illustrates an example 3D context. We estimate $p(v_i|v_{i-1}, \ldots, v_1)$ using a neural network which we dub \textbf{VoxelDNN}.  

\par The conditional distributions in~\eqref{eq:p(v)} depend on previously decoded voxels. This requires a \textit{causality} constraint on the VoxelDNN network. To enforce causality, we extend to 3D the idea of masked convolutional filters, initially proposed in PixelCNN~\cite{oord2016pixel}. Specifically, two kinds of masks (A or B) can be employed. Type A mask is filled by zeros from the center position to the last position in raster scan order as shown in Figure \ref{fig:context}(b). Type B mask differs from type A in that the value in the center location is 1. We apply type A mask to the first convolutional layer to restrict both the connections from all future voxels and the voxel currently being predicted.  In contrast, from the second convolutional layer, type B masks are applied which relaxes the restrictions of mask A by allowing the connection from the current spatial location to itself. 
%filled by zeros from the center position to the last position in raster scan order, 
%The examples of 3D type A and B masks are shown in Figure \ref{fig:maskab}.
%Similarly to PixelCNN~\cite{oord2016pixel}, we apply a type A mask to the input convolution and type B mask to purely convolutional layers to enforce causality.

%The value in centre position in mask B is `1'.  %The masked filters in \cite{mentzer2018conditional} , \cite{salimans2017pixelcnn++} have been shown as an effective solution to solve the problem. 
%For each input block of size $d \times d \times d$, VoxelDNN predicts the distribution of each voxel $v_i$ given the previously scanned voxels. During training,

%    L&= \E_{v \sim p(v)} \left [  \right ]

\par In order to learn good estimates $\hat{p}(v_i|v_{i-1}, \ldots, v_1)$ of the underlying voxel occupancy distribution $p(v_i|v_{i-1}, \ldots, v_1)$, and thus minimize the coding bitrate, we train VoxelDNN using cross-entropy loss. That is, for a block $v$ of resolution $d$, we minimize :
\begin{equation}\label{eq:CEloss}
    H(p,\hat{p}) = \mathbb{E}_{v\sim p(v)}\left[\sum_{i=1}^{d^3} -\log \hat{p}(v_i)\right].
\end{equation}
It is well known that cross entropy represents the extra bitrate cost to pay when the approximate distribution $\hat{p}$ is used instead of the true $p$. More precisely, $H(p,\hat{p}) = H(p) + D_{KL}(p\| \hat{p})$, where $D_{KL}$ denotes the Kullback-Leibler divergence and $H(p)$ is Shannon entropy. Hence, by minimizing~\eqref{eq:CEloss}, we indirectly minimize the distance between the estimated conditional distributions and the real data distribution, yielding accurate contexts for arithmetic coding. Note that this is different from what is typically done in learning-based \textit{lossy} PC geometry compression, where the focal loss is used~\cite{quach2019learning, quach2020improved}. The motivation behind using focal loss is to cope with the high spatial unbalance between occupied and non-occupied voxels. The reconstructed PC is then obtained by hard thresholding $\hat{p}(v)$, and the target is thus the final classification accuracy. Conversely, here we aim at estimating accurate soft probabilities to be fed into an arithmetic coder.

\vspace{-6mm}

% VoxelDNN aims to minimize the Kullback–Leibler divergence (KL divergence) between the ground truth distribution $p$ of the voxel and the predicted distribution $q$. Cross entropy (CE) is the sum of the entropy of the voxel occupation knowing the true distribution plus the KL divergence. Minimizing KL divergence is equivalent to minimizing the cross entropy. Our final loss $L$ is the average of losses on all $v_i$ (Equation \ref{eq:loss}). In contrast to other deep approaches for lossy PC geometry compression, we estimate soft information instead of using focal loss \cite{quach2019learning,quach2020improved} and hard thresholding probabilities.

% \begin{equation}
% \begin{split}
%     L&=\frac{1}{d^{3}} \sum^{d^{3}}_{i=1}CE_{i}\\
%     &=\frac{1}{d^{3}} \sum^{d^{3}}_{i=1}-v_i\cdot \log(q_i)-(1-v_i)\cdot \log(1-q_i)
% \end{split}
% \label{eq:loss}
% \end{equation}

%\begin{figure}
%\centering
%\begin{minipage}{.5\linewidth}
%  \centering
%  \includegraphics[width=0.8\linewidth]{figures_pcc/masktypeA1.png}
%\end{minipage}%
%\begin{minipage}{.5\linewidth}
%  \centering
%  \includegraphics[width=0.8\linewidth]{figures_pcc/masktypeB1.png}
%\end{minipage}
%\caption{Examples of $3 \times 3 \times 3$ 3D masks. Left: type A mask, right: type B mask. The difference is the value in the center.}
%\label{fig:maskab}
%\end{figure}

\input{accBLockResult}

\par Figure \ref{fig:Networkarchitecture} shows our VoxelDNN network architecture. Given the $64 \times 64 \times 64$ input block, VoxelDNN outputs the predicted occupancy probability of all input voxels. Our first 3D convolutional layer uses $7 \times 7 \times 7$ kernels with a type A mask. Type B masks are used in the subsequent layers. To avoid vanishing gradients and speed up the convergence, we implement two residual connections with $5 \times 5 \times 5$ kernels.  Throughout VoxelDNN, the ReLu activation function is applied after each convolutional layer, except in the last layer where we use softmax activation. In total, our model contains 290,754 parameters and requires less then 4MB of disk storage space.
%\input{encoder_algrth}
%\vspace{-1.2cm}

\subsection{Multi-resolution encoder and adaptive partitioning}\label{ssec:multires}

We use an arithmetic coder to encode the voxels sequentially from the first voxel to the last voxel of each block in a generative manner. Specifically, every time a voxel is encoded, it is fed back into VoxelDNN to predict the probability of the next voxel. Then, we pass the probability to the arithmetic coder to encode the next symbol. 

\begin{figure}
\captionsetup{singlelinecheck = false, format= hang, justification=raggedright, font=small, labelsep=space}
\begin{minipage}[b]{.24\linewidth}
  \centering
  \centerline{\includegraphics[width=0.90\linewidth]{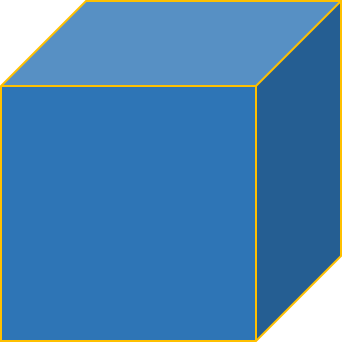}}
%  \vspace{1.5cm}
  \centerline{(a)}\medskip
\end{minipage}
\hfill
\begin{minipage}[b]{0.24\linewidth}
  \centering
  \centerline{\includegraphics[width=0.90\linewidth]{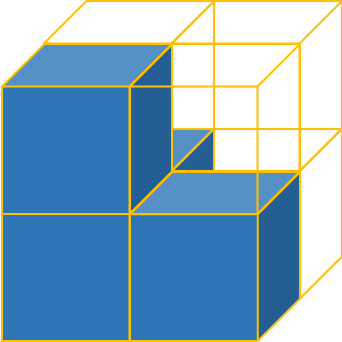}}
%  \vspace{1.5cm}
  \centerline{(b) }\medskip
\end{minipage}
\hfill
\begin{minipage}[b]{0.24\linewidth}
  \centering
  \centerline{\includegraphics[width=0.90\linewidth]{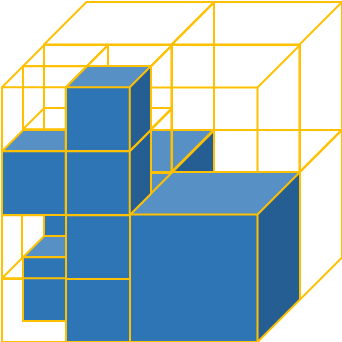}}
%  \vspace{1.5cm}
  \centerline{(c) }\medskip
\end{minipage}
\hfill
\begin{minipage}[b]{0.24\linewidth}
  \centering
  \centerline{\includegraphics[width=0.90\linewidth]{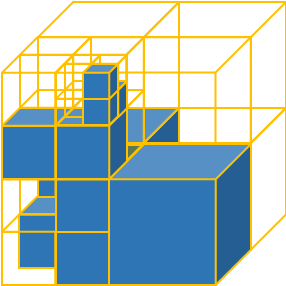}}
%  \vspace{1.5cm}
  \centerline{(d)}\medskip
\end{minipage}
\caption{Partitioning a block of size 64 into: (a) a single block of size 64, (b): blocks of size 32, (c): 32 and 16, (d): 32, 16 and 8. Non-empty blocks are indicated by blue cubes.}
\label{fig:4levelpartiitoning}
\end{figure}

However, applying this coding process at a fixed resolution $d$ (in particular, on blocks 64), can be inefficient when blocks are sparse, i.e.,  they contain only few occupied voxels. 
% encoding the whole block as a single block is not always the best solution, especially when blocks are sparse. 
This is due to the fact that in this case, there is little or no information available in the receptive fields of the convolutional filters. To overcome this problem, we propose a rate-optimized multi-resolution splitting algorithm as follows.
We partition a block into 8 sub-blocks recursively and signal the occupancy of sub-blocks as well as the partitioning decision (0: empty, 1: encode as single block, 2: further partition). The partitioning decision depends on the output bits after arithmetic coding. If the total bitstream of partitioning flags and occupied sub-blocks is larger than encoding parent block as a single block, we do not perform partitioning. The details of this  process are shown in Algorithm \ref{algo:proposed_method}. The maximum partitioning level is controlled by $maxLv$ and  partitioning is performed up to $maxLv=5$ corresponding to a smallest block size of 4. Depending on the output bits of each partitioning solution, a block of size 64 can contain a combination of blocks with different sizes. Figure \ref{fig:4levelpartiitoning} shows 4 partitioning examples for an encoder with $maxLv=4$.
Note that VoxelDNN learns to predict the distribution of the current voxel from previous encoded voxels. As a result, we only need to train a \textit{single} model to predict the probabilities for different input block sizes. 
%In particular, VoxelDNN can perform prediction on blocks of size $d \times d \times d$ with $d$ ranging from 2 to 64. 
% This enables the encoder to decide whether to encode the block as a single block or to partition it into smaller blocks, flag out empty child blocks and only perform prediction on occupied blocks.

\setlength{\textfloatsep}{20pt}

%% file: encoder_algrth.tex
\begin{algorithm2e}[t]
\footnotesize
\SetAlgoLined
\SetKwInOut{Input}{Input}
\SetKwInOut{Output}{Output}
\KwIn{block: $B$, current level: $curLv$, max level: $maxLv$} 
\KwOut{partitioning flags: $fl$, output bitstream: $bits$}
\SetKwFunction{FMain}{partitioner}
\SetKwProg{Fn}{Function}{:}{}
\Fn{\FMain{$B, curLv, maxLv$}}{
    %\tcc{encode as 8 child blocks}
    $fl2$ $\leftarrow$ 2  \tcp*[l]{encode as 8 child blocks}
    \For{block $b$ in child blocks of $B$}{
        \eIf{$b$ is empty}{
            $child\_flag$ $\leftarrow$ 0\;
            $child\_bit$ $\leftarrow empty$\;
        }{
            \eIf{$curLv$ $==$ $maxLv$}{
                %\tcc{encode child block as a single block}
                $child\_flag$ $\leftarrow$ 1\;
                $child\_bit$ $\leftarrow$ singleBlockCoder($b$)\;
            }{
                %\tcc{paritioning selection at lower level}
                $child\_flag, child\_bit$ $\leftarrow$ partitioner($b$, $curLv+1$,$maxLv$)\;
            }
        }
        $ fl2 \leftarrow$ [$fl2$, $child\_flag$]\;
        $ bit2\leftarrow$ [$bit2$, $child\_bit$]\; 
    }
    $total\_bit2=sizeOf(bit2)+len(fl2)\times 2$\;
    %\tcc{encode as a single block}
    $fl1 \leftarrow$ 1\tcp*[l]{encode as a single block}
    $bit1 \leftarrow$ singleBlockCoder($B$)\;
    $total\_bit1=sizeOf(bit1)+len(fl1)\times 2$\;
    \tcc{partitioning selection}
    \eIf{$total\_bit2 \geq total\_bit1$}{
        \KwRet $fl1, bit1$\;
    }{
        \KwRet $fl2, bit2$\;
    }
  }
\caption{Block partitioning selection}
\label{algo:proposed_method}
\end{algorithm2e}
%\setlength{\textfloatsep}{0pt}

%% file: accBLockResult.tex
\begin{center}
\centering
\begin{table*}[ht]
\caption{Average rate in bpov of VoxelDNN at different partitioning levels compared with MPEG G-PCC v12 and P(PNI).}
\resizebox{0.97\linewidth}{!}{ \begin{tabular}{|P{1.5cm}|l||R{1.2cm}|R{1.2cm}||R{1.2cm}|R{1.5cm}|R{1.2cm}|R{1.5cm}|R{1.2cm}|R{1.5cm}|R{1.2cm}|R{1.5cm}|}
\cline{3-12}
% \multicolumn{2}{|c||}{\begin{bf} Test PC \end{bf}}
\multicolumn{2}{c|}{}
& \multicolumn{1}{|c|}{\begin{bf} P(PNI) \end{bf}}
& \multicolumn{1}{c||}{\begin{bf} G-PCC \end{bf}}
& \multicolumn{2}{c|}{\begin{bf} block 64 \end{bf}}
& \multicolumn{2}{c|}{\begin{bf} block 64 + 32 \end{bf}}
& \multicolumn{2}{c|}{\begin{bf} block 64 + 32 + 16 \end{bf}}
& \multicolumn{2}{c|}{\begin{bf} block 64 + 32 + 16 + 8 \end{bf}}
\\
\hline
Dataset&Point Cloud &bpov& bpov&bpov&Gain over G-PCC &bpov&Gain over G-PCC &bpov&Gain over G-PCC&bpov&Gain over G-PCC\\
\hline
\multirow{5}{*}{MVUB}&Phil9 &1.88& 1.2285&0.9819 &20.07\% &0.9317 &24.16\% &0.9203 &25.09\% &  0.9201&25.10\%\\
\cline{2-12}
%\hline
&Ricardo9 &1.79&1.0423 & 0.7910& 24.11\%&0.7276 & 30.19\%& 0.7175& 31.16\%& 0.7173&31.18\% \\
\cline{2-12}
&Phil10&- & 1.1599 & 0.8941&22.92\% &0.8381 & 27.74\%&0.8308 & 28.37\%&  0.8307&28.38\% \\
\cline{2-12}
&Ricardo10&- & 1.0673 & 0.8108&24.03\% &  0.7596&28.83\% &0.7539&29.36\% &  0.7533&29.42\% \\
\cline{2-12}
&\textbf{Average}&\textbf{1.84}  &\textbf{1.1245}  &\textbf{0.8695}&\textbf{22.78\%} &\textbf{0.8143 }&\textbf{27.73\%} &\textbf{0.8056 }&\textbf{28.50\%}& \textbf{0.8054}&\textbf{28.52\% } \\
\cline{2-12}

\hline
\multirow{5}{*}{MPEG 8i}&Loot10 &1.69& 0.9525 &0.7016 & 26.34\%&0.6464 &32.14\% &0.6400 & 32.81\% &0.6387 &32.94\%\\
\cline{2-12}
%\hline
&Redandblack10 &1.84&1.0890  & 0.7921&27.26\% &0.7383 &32.20\% & 0.7317&32.81\%& 0.7317&32.81\% \\
\cline{2-12}
&Boxer9&- &1.0816  & 0.8034&25.72\% &0.7620 & 29.55\%&0.7558 &30.12\% &0.7560 &30.10\% \\
\cline{2-12}
&Thaidancer9&- &1.0679 &0.8574 &19.71\% &0.8145 &23.73\% &0.8091 &24.23\%&0.8078   &24.36\%\\
\cline{2-12}

&\textbf{Average}& \textbf{1.77} &\textbf{1.0478} &\textbf{0.7886} &\textbf{24.76\%} &\textbf{0.7403} &\textbf{29.40\% }&\textbf{0.7342} &\textbf{29.99\% }& \textbf{ 0.7336}&\textbf{30.05\% }\\
\hline
\hline
\end{tabular}}
\label{table:addingblocksize}

\end{table*}
\end{center}
%\vspace{-1cm}
\setlength{\textfloatsep}{5pt}% Remove \textfloatsep

%% file: perform.tex
\subsection{Experimental Setup}
\textbf{Training dataset:} Our models were trained on a combined dataset from ModelNet40 \cite{wu20153d}, MVUB \cite{loop2016microsoft} and 8i \cite{d20178i,8i} datasets. We sample 17 PCs from Andrew10, David10, Sarah10 sequences in MVUB dataset into the training set. In 8i dataset, 18 PCs sampled from Soldier and Longdress sequence are selected. The ModelNet40 dataset, is  sampled  into depth 9 resolution and  the 200 largest PCs are selected. Finally, we divide all selected PCs into occupied blocks of size $64\times64\times64$. In total, our dataset contains 20,264 blocks including 4,297 blocks from 8i, 4,820 blocks from MVUB and 11,147 blocks from ModelNet40. We split the dataset into 18,291 blocks for training and 1,973 blocks for testing.\\
\textbf{Training:} VoxelDNN is trained with Adam \cite{adam} optimizer, a learning rate of 0.001, and a batch size of 8 for 50 epochs on a GeForce RTX 2080 GPU.\\
\textbf{Evaluation procedure:} We evaluate our methods on both 9 and 10 bits depth PCs that have the lowest and highest rate performance when testing with G-PCC method from the MVUB and MPEG datasets. These PCs were not used during training. The effectiveness of the partitioning scheme is evaluated by  increasing the maximum partitioning level from 1 to 5 corresponding to  block sizes  64, 32, 16, 8 and 4.
%Figure \ref{table:exsetup} summaries the parameters in our experiments.

%\begin{center}
%\centering
%\begin{table}[H]
%\caption{Experiment setup}
%\resizebox{\linewidth}{!}{\begin{tabular}{|M{2cm}|P{3cm}|P{7.5cm}|}
%\hline
%\begin{bf} Models \end{bf} &\begin{bf} Parameter \end{bf} & \begin{bf} Value  \end{bf}  \\
%\hline
%\multirow{7}{*}{VoxelDNN}&Training set& Mixed of ModelNet 40, Microsoft Voxelized Upper Bodies and 8i Voxelized Full Bodies \\
%\cline{2-3}
% & Test set & 8i  Voxelized  Full  Bodies, CAT1 (MPEG),  Microsoft  Vox-elized Upper Bodies\\
%\cline{2-3}
% & Epochs & 50\\
% \cline{2-3}
%  & Batch size & 8\\
%\hline
%\end{tabular}}
%\label{table:exsetup}
%\end{table}
%\end{center}
%)àç!è
  
%\vspace{-1.5cm}

\subsection{Experimental results}
%\subsubsection{Training results}
%\par Our VoxelDNN predicts the occupancy probabilities of each voxels. In addition to Cross Entropy loss to evaluate the performance during training, voxels are classified as empty or occupied using a threshold of 0.5. Then, the predicted occupancy is compared with the ground truth occupancy to obtain binary classification metrics. Table \ref{table:VoxelDNNtrainresult} illustrates the training results by 5 metrics: accuracy, precision, recall, specificity and F1 score. Since the point clouds are very sparse and most of the voxels are empty, the model achieve 99\% on accuracy in just few epochs. We observed that F1 score, Precision, and Recall are effective metrics to evaluate the performance of our VoxelDNN model. We stopped training after 50 epochs as the monitored quantities have stopped improving.
%\begin{center}
%\centering
%\begin{table}[H]
%\caption{VoxelDNN training results}
%\resizebox{\linewidth}{!}{\begin{tabular}{|M{2cm}|P{2cm}|P{2cm}|P{1.6cm}|P{2cm}|P{2cm}|}
%\hline
%\begin{bf} Metric \end{bf} & \begin{bf} Accuracy  \end{bf} &\begin{bf} Precision   \end{bf}& \begin{bf} Recall   \end{bf} & \begin{bf} Specificity   \end{bf} & \begin{bf} F1 score   \end{bf} \\
%\hline
%VoxelDNN &0.99&0.81&0.760&0.99&0.79\\
%\hline
%\end{tabular}}
%\label{table:VoxelDNNtrainresult}
%\end{table}
%\end{center}
%\vspace{-1.5cm}
\input{voxelDNNResults}

%% file: voxelDNNresults.tex
%\subsubsection{VoxelDNN}
\par Table \ref{table:addingblocksize} reports the average rate in bits per occupied voxel (bpov) of the proposed method for 4 partitioning levels, compared with G-PCC v.12~\cite{graziosi2020overview}. We also report the results of the recent intra-frame geometry coding method P(PNI) from \cite{garcia2019geometry} (the coding results are available only for some of the tested PCs). The results with 5 partitioning levels are identical to 4 partitioning levels and are not shown in the table. In all  experiments, the total size of signaling bits for the high-level octree and partitioning accounts for less than $2\%$ of the bitstream.
\par
\begin{figure}

\centering
\captionsetup{justification=centering}
\captionsetup{singlelinecheck = false, format= hang, justification=raggedright, font=small, labelsep=space}
\includegraphics[width=\linewidth]{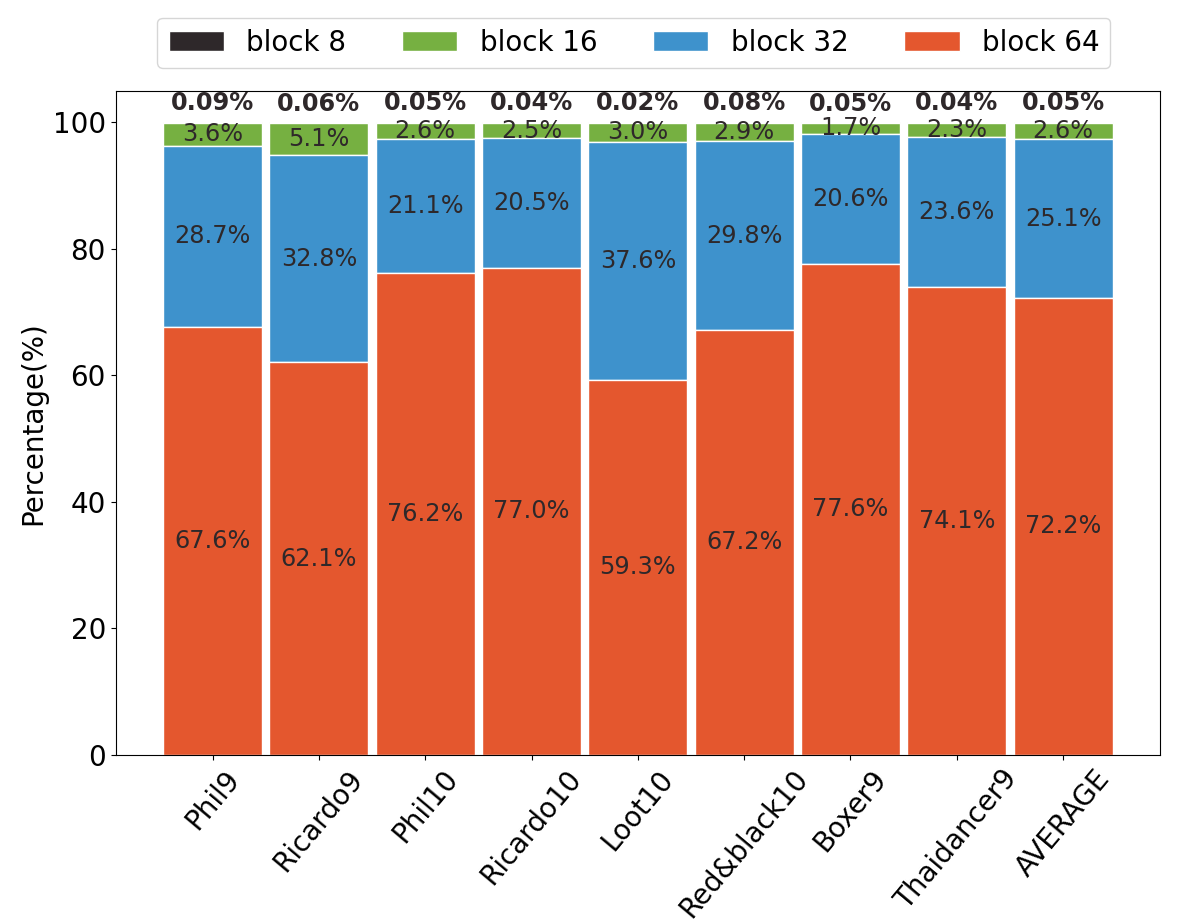}
\caption{Percentage of encoded points in each block size. From top to bottom: block 8, 16, 32, 64.}
\label{fig:points_on_each_block}
\end{figure}

 We observe that our proposed solution at all 4 levels and G-PCC outperform P(PNI) by a large margin. VoxelDNN solutions outperform G-PCC on the MVUB and MPEG 8i datasets with the highest average rate saving of $28.4\%$ and $29.9\%$, respectively.
As partitioning levels increases, the corresponding gain over G-PCC also increases; however, there is only a slight increase with 3 and 4 levels compared to the gain of the lower level. 
This can be explained with Figure \ref{fig:points_on_each_block}, which shows the percentage allocation of the encoded points in each partitioning size, for different PCs, after optimal partitioning.
% the percentages of the total occupied voxels/points for each partition size in different PCs. 
It can be seen that most voxels are encoded using blocks $64$ and $32$, while very few ones are encoded with blocks of smaller size. 
While adding more partition levels enables to better adapt to point cloud geometry, smaller partitions entail a higher signalization cost. This is not often compensated by a bitrate reduction of the  sub-blocks, since in the smaller partitions the encoder has access to limited contexts, resulting in less accurate probability estimations.

%table block size percentage

%% file: conclusion.tex
\par This paper  proposed a hybrid octree/voxel-based lossless compression method for point cloud geometry. It employs for the first time a deep generative model in the voxel space to estimate the  occupancy probabilities sequentially.
% VoxelDNN method to losslessly encode the point cloud geometry in a generative manner while the previously encoded voxels are used to predict the distribution of the current voxel. 
Combined with a rate-optimized partitioning strategy, the proposed method outperforms MPEG G-PCC with average 28\% rate savings over all tested datasets.
% Besides, VoxelDNN offers multi-resolution block coding to adapt to the local point cloud characteristics. Our results show that the proposed multi-resolution VoxelDNN outperforms MPEG G-PCC with average 28\% rate savings over all test datasets. 
% In future work, we plan to extend our approach to jointly encode geometry and attribute information, thus allowing an end-to-end point cloud coding method.
We are now working on improving lossless coding of PC geometry by using more powerful generative models, and jointly optimizing octree and voxel coding. 
%This paper presents a deep learning-based point cloud geometry compression method. W